\documentclass{article}

\usepackage{arxiv}
\usepackage{ifsym}
\usepackage{times}
\usepackage[ruled,vlined]{algorithm2e}
\usepackage{latexsym}
\usepackage{xurl}
\usepackage{hyperref}
\usepackage{url}
\usepackage{paralist}
\usepackage{xcolor}
\usepackage{colortbl}
\usepackage{multirow}
\usepackage{mathtools}
\usepackage[export]{adjustbox}
\usepackage{amsmath}
\DeclareMathOperator*{\argmax}{arg\,max}

\usepackage{subfig}

\newcommand\Tstrut{\rule{0pt}{2.1ex}}       
\newcommand\Bstrut{\rule[-0.9ex]{0pt}{0pt}} 

\usepackage{scalerel,stackengine}

\usepackage{bbm}

\usepackage{microtype}

\title{`Beach' to `Bitch': Inadvertent Unsafe Transcription of Kids' Content on YouTube}

\author{
Krithika Ramesh\thanks{Krithika Ramesh was equally advised by Ashiqur R. KhudaBukhsh and Sumeet Kumar. This work was performed at Indian School of Business.} \\
  \small{Manipal University} \\
  \texttt{kramesh.tlw@gmail.com} \\
  \And
Ashiqur R. KhudaBukhsh\thanks{Ashiqur R. KhudaBukhsh and Sumeet Kumar are joint corresponding authors.}  \\
  \small{Rochester Institute of Technology}\\
  \texttt{axkvse@rit.edu} \\
 \And
Sumeet Kumar$^\dagger$\\
  \small{Indian School of Business}\textsuperscript{\rm 2}\\
  \texttt{Sumeet\_Kumar@isb.edu } \\
}

\begin{document}
\maketitle

\begin{abstract}

Over the last few years, YouTube Kids has emerged as one of the highly competitive alternatives to television for children's entertainment. Consequently, YouTube Kids' content should receive an additional level of scrutiny to ensure children's safety. While research on detecting offensive or inappropriate content for kids is gaining momentum, little or no current work exists that investigates to what extent AI applications can (accidentally) introduce content that is inappropriate for kids. 

In this paper, we present a novel (and troubling) finding that well-known automatic speech recognition (ASR) systems may produce text content highly inappropriate for kids while transcribing YouTube Kids' videos. We dub this phenomenon as \emph{inappropriate content hallucination}. Our analyses suggest that such hallucinations are far from occasional, and the ASR systems often produce them with high confidence. We release a first-of-its-kind data set of audios for which the existing state-of-the-art ASR systems hallucinate inappropriate content for kids. In addition, we demonstrate that some of these errors can be fixed using language models.

\end{abstract}

\keywords{Online Kids Content \and  ASR  \and Inappropriate Content Hallucination}

\section{Introduction}

Over the last few years, YouTube Kids has emerged as one of the highly competitive alternatives to television for children's entertainment. For example, between 2015 to 2021, the subscriber count of Ryan's World, a highly popular YouTube channel for kids, rose from 32K to 30.3 millions\footnote{Source: \url{https://web.archive.org/}}. With this steep viewership growth, content hosted in these channels has received escalated scrutiny. Following recent reports indicating occasional slip-ups in YouTube Kids' content moderation systems, recent works have explored automatic detection of inappropriate content from videos to aid human moderation ~\cite{papadamou2020disturbed,han2020discovery, alghowinem2018safer}.

\begin{figure}[htb]
\centering
\includegraphics[trim={0 0 0 0},clip, height=2.0in]{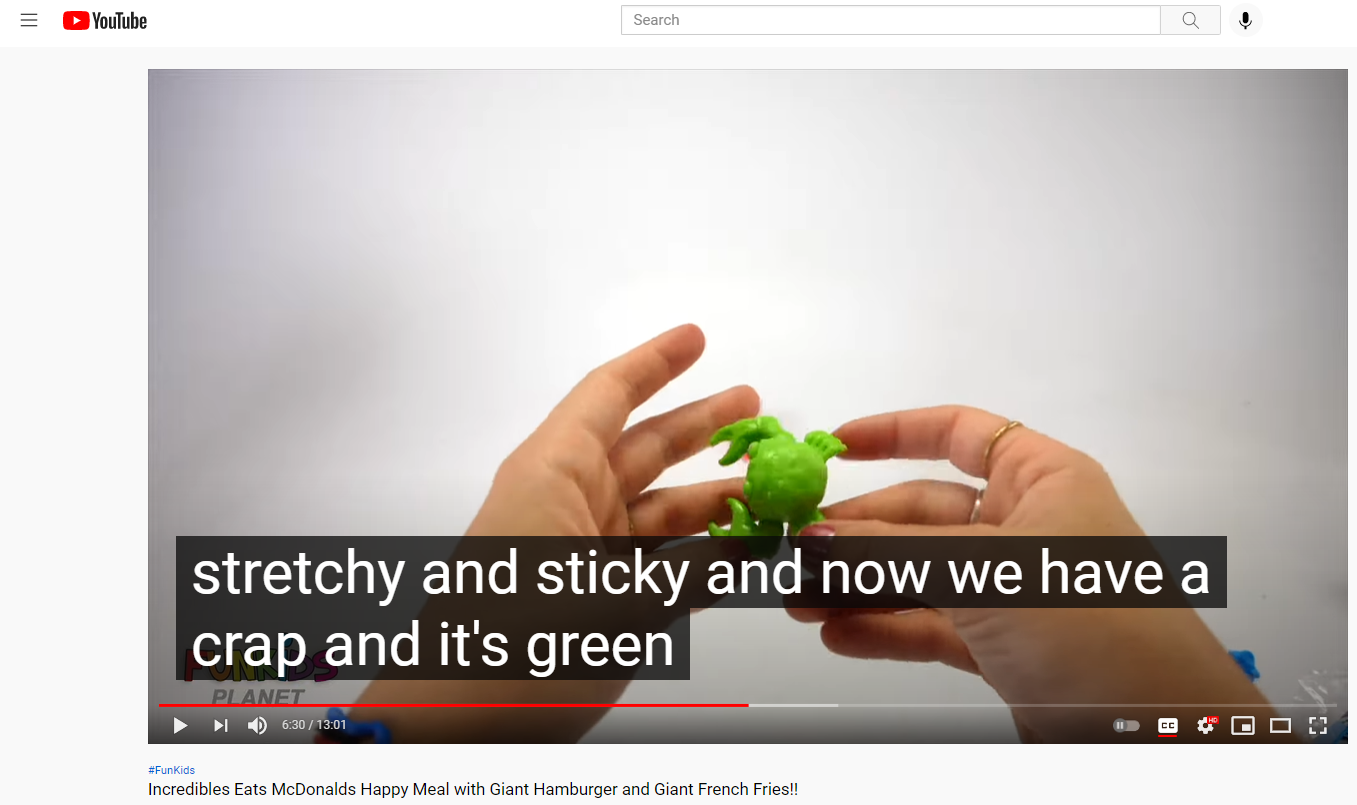}
\caption{Example of \emph{inappropriate content hallucination}. A screenshot of a YouTube video taken from a popular channel \emph{Fun Kids Planet} with 1.34m subscribers. The ASR (automatic speech transcription) system incorrectly transcribes \texttt{crab} as \textbf{\textcolor{red}{\texttt{crap}}}. } 
\label{fig:Crap}
\end{figure}

\renewcommand{\tabcolsep}{2mm}
\begin{table*}[t]
{
\small
\begin{center}
     \begin{tabular}{|p{0.4\textwidth}|p{0.4\textwidth}|}
  \hline
  Ground truth & Transcription \\
    \hline

\cellcolor{blue!15}\ldots its fluffy and so so soft toy\ldots & \cellcolor{red!15}
\ldots its 
\textbf{\textcolor{red}{fucking }} and so so soft toy\ldots \\
   \hline 
\cellcolor{blue!15} \ldots if you like this craft keep on watching until the end so you can see related videos\ldots & \cellcolor{red!15} \ldots if you like this \textbf{\textcolor{red}{crap}} keep on watching until the end so you can see related videos\ldots \\
   \hline    
\cellcolor{blue!15} \ldots stretchy and sticky and now we have a crab and its green\ldots & \cellcolor{red!15} \ldots stretchy and sticky and now we have a \textbf{\textcolor{red}{crap}} and its green\ldots \\
   \hline    
   
\cellcolor{blue!15} \ldots duck pulls away here comes arthur mr conductor tries as hard as duck gets a little bit of boost but its not enough\ldots & \cellcolor{red!15} \ldots duck pulls away here comes arthur mr conductor tries as hard as \textbf{\textcolor{red}{fuck}} gets a little bit of boost but its not enough\ldots \\
   \hline
\cellcolor{blue!15} \ldots in order to be strong and brave like heracles\ldots & \cellcolor{red!15} \ldots in order to be strong and  \textbf{\textcolor{red}{rape}} like heracles\ldots \\
   \hline

    \end{tabular}
    
\end{center}
\caption{{Examples where ASR systems hallucinated taboo-words originally not present in the source.}}
\label{tab:example}}
\end{table*}

While detecting offensive or inappropriate content for specific demographics is a well-studied problem, such studies typically focus on detecting offensive content present in the source, not how objectionable content can be (accidentally) introduced by a downstream AI application. With the growing trend of complex AI pipelines where one sub-system's output is another sub-system's input, it is increasingly becoming more important to acknowledge that inappropriate content may not always be present in the source, it can inadvertently creep in through a downstream AI application. We call this phenomenon as \emph{inappropriate content hallucination}. In this paper\footnote{This paper has been accepted at AAAI 2022, AI for Social Impact track.}, we study the interaction between YouTube Kids video content and ASR (automatic speech recognition) systems to assess this phenomenon. ASR systems have prominent use in multimedia systems to generate transcripts for video content at scale and have found use in a broad range of applications that include call routing~\cite{riccardi1997spoken}, transcribing meetings and lectures~\cite{ranchal2013using}, IoT appliances~\cite{mehrabani2015personalized}, and medical scribing~\cite{finley2018automated}. Our study uncovers a disturbing pattern of (inadvertent) introduction of inappropriate words by ASR systems while transcribing kids' videos described through the following illustrative example.   

\subsection{An Illustrative Example}

Consider the  real-world example from a highly popular YouTube Kids channel presented in Figure~\ref{fig:Crap}. All words present in the utterance are correctly transcribed except for the word \texttt{crab}, which is incorrectly transcribed as \texttt{crap}. Among the broad taxonomy of offensive/inappropriate words such as profanity, slur, insults, and slang, following developmental psychology literature on children~\cite{jay1992cursing}, we broadly categorize such inappropriate words as \emph{taboo-words}.

We do not know what percentage of kids use YouTube Kids application (as opposed to YouTube) and what fraction of kids watching videos turn their subtitles on while watching the videos. However, documented evidence exists indicating that (1) \emph{same language subtitling}~\cite{vanderplank2016effects} and captioned media in foreign language~\cite{vanderplank2016captioned} both improve learning in children; and (2) for children with disabilities, captions provide a critical learning resource~\cite{vanderplank2016captioned}. In the example presented in Figure~\ref{fig:Crap} where every other word in the audio matches with the transcript but the taboo-word, it is not hard to envision that there is a potential risk for kids to incorporate this taboo-word into their vocabulary without even knowing that it is inappropriate.

As indicated in Table~\ref{tab:example}, such hallucinations of taboo-words are far from a one-off incident and some of the videos containing highly inappropriate hallucinated taboo-words could have been exposed to millions of viewers. Of course, ASR for kids adds additional challenges and new methods have investigated possible mitigation strategies~\cite{wu2019advances, yeung2018difficulties,plantinga2019towards}. However, to our knowledge, no such study considered this novel risk of introducing unsafe content through transcription errors. Our study thus adds value to the ongoing conversation around the potential risks and harms to the society that can be caused by over-reliance on AI systems and applying them to a  demographic potentially underrepresented in the training data (in our case, kids). 

Further, one key point to remember is that none of these taboo-words was present in the actual content -- they are (accidentally) introduced by a downstream AI application. As already mentioned, with complex and connected AI applications where outputs of one AI application may form inputs to others\footnote{In fact, existing multimodal approaches~\cite{alghowinem2018safer} have consulted ASR outputs to detect inappropriate content in videos.}, these inadvertently generated taboo-words can pose unseen challenges for downstream applications.

While our first goal in this paper is to attract the attention of the research community to this novel threat of accidental introduction of inappropriate content through downstream AI applications, in order to facilitate robust testing of ASR systems, one of our key contributions is a novel data set of challenging audio inputs in which major ASR systems hallucinated taboo-words\footnote{Data set, lexicon, and additional details are available at our \href{https://github.com/sumeetkr/UnsafeTranscriptionofKidsContent}{project page}: \url{https://github.com/sumeetkr/UnsafeTranscriptionofKidsContent}.}. In addition, we demonstrate that some of these challenging instances can be corrected using high-performance language models.

\textbf{Contributions:} Our contributions are the following.
\begin{compactenum}
\item \emph{Social:} Via a comprehensive study of 7,103 videos on 24 YouTube Kids (YTK) channels, our study indicates that prominent ASR systems often hallucinate taboo-words in children's content. 
\item \emph{Resource:} We release a first-of-its-kind data set of 652 challenging audio inputs in which prominent ASR systems have hallucinated taboo-words along with the ground truth transcriptions.  We release a lexicon of 1,301 taboo-words for kids that draws from developmental psychology literature, a curated corpus MPAA rated movie subtitles data set and a well-known hate lexicon.
\item\emph{Method:} We present a novel application of masked language models to fix some of these errors.
\end{compactenum}

\section{Related Work}

Although transcripts generated using ASR systems lack in terms of quality as compared to manual transcription performed by human annotators, ASR systems have found use in a broad range of applications that include call routing~\cite{riccardi1997spoken}, transcribing meetings and lectures~\cite{ranchal2013using}, video subtitling~\cite{sawaf2012automatic}, IoT appliances~\cite{mehrabani2015personalized}, and medical scribing~\cite{finley2018automated}. In addition to challenges such as background noise~\cite{rajnoha2011asr} and speech variability~\cite{benzeghiba2007automatic}, and specifically while transcribing kids' speeches~\cite{wu2019advances, yeung2018difficulties,plantinga2019towards}, ASR systems face issues with comprehending speech where stuttering, erratic pauses, dysarthric speech, etc. are present \cite{10.5555/2018192.2018228}.  

Given the disparity in the latency between manual transcription and ASR transcripts, \cite{10.1145/2899475.2899478} propose allowing the transcriptionists to utilize the ASR outputs as a starting point for their transcription work, as manual transcription would otherwise take up to almost five times the length of the audio. It was noted that this did indeed reduce the normalized average latency, provided that the WER (Word Error Rate) of the ASR system was less than 30\%.

Other approaches to improve the quality of the audio transcripts involve grammatical correction. Several lines of work exist that have used language models to improve ASR systems~(see, e.g., \cite{arisoy2015bidirectional,chen2017investigating,shin2019effective,DBLP:journals/corr/abs-2103-14580}).  \cite{DBLP:journals/corr/abs-2103-14580} introduce WLM-SC, a specific variant of warped language models, a generalized version of masked language models trained on data sets containing grammatical errors so as to become robust to these word-level errors. WLM-SC is used for sentence correction and not only implements the warping operation across the tokens in the sentence, but also indicates if a warping position took place at each position, and the model is trained to predict the token as well as the warping operation used. It is demonstrated that WLM-SC not only improves the WER of automatic transcriptions but that of human transcriptions as well.

\cite{kim2019comparison} draw a comparison between multiple ASR systems, and manual transcription ones as well, which outstrip the automatic ones in terms of their WER. In addition, they also attempt to find a correlation between nonverbal behavior cues and unintelligible speech, showing that the variability of the speech intensity is lower when the speech is not clear. Our work contrasts with existing research on ASR systems in two key ways: (1) unlike traditional performance metrics like word error rate (WER), we focus on a potentially harmful content hallucination of ASR systems, a phenomenon never studied before to our knowledge; and (2) we release a novel benchmark data set of challenging hallucinated examples with ground truth.  

At a philosophical level, our work is closely related to 
\cite{10.1145/3442188.3445922, DBLP:conf/emnlp/GehmanGSCS20} that explore the potential risks associated with large opaque models, including the lack of diversity in the data they are trained on, and the biases they exhibit. In particular, the work discusses how biases can reflect in the derogatory language that could potentially be produced by the model, in the form of racial slurs and derogatory terms that target marginalized communities, and the difficulty of filtering out such terms from our data.

\begin{center}
\begin{table}[ht!]
\small
\centering
 \begin{tabular}{|p{3.4cm}|p{2cm}|p{2cm}|p{2cm}|p{2cm}|p{2cm}|} 
 \hline
 Channel Name & Channel View Count & Channel Subscriber Count  & \# of Videos & \# of Google Transcripts with taboo-words & \# of AWS Transcripts with taboo-words  \\ [0.5ex] 
 \hline
 \hline
 
Sesame Street&20 Billion&23 Million&2405&432 (2)&763 (122)\\\hline
Ryan's World&50 Billion&32 Million&1437&892 (9)&1,138 (383)\\\hline
3KidsTV&2 Million&10 Thousand&39& 8 (0)&12 (0)\\\hline
Sesame Studios&343 Million&615 Thousand&320&62(0)&113(10)\\\hline
Barbie&3 Billion&11 Million&98&26 (1)&32 (6)\\\hline
Moonbug Kids - Cartoons \& Toys &110 Million&3 Million&632&497(31)&499(63)\\\hline
Elizabeth \& Eva TV&42 Million&172 Thousand&157&57 (1)&87 (30)\\\hline
Rob The Robot - Learning Videos For Children&215 Million&380 Thousand&113&90 (3)&103 (24)\\\hline
SimpleCrafts - 5 Minute Crafts For All&211 Million&718 Thousand&736&257 (1)&333 (16)\\\hline
Kids Toys Play&524 Million&461 Thousand&112&76 (1)&91 (22)\\\hline
Mister Max&14 Billion&22 Million&72&8 (0)&31 (12)\\\hline
Blippi - Educational Videos for Kids&12 Billion&15 Million&78&57 (0)&64 (15)\\\hline
Like Nastya&69 Billion&86 Million&224&44 (1)&76 (31)\\\hline
New Sky Kids&2 Billion&2 Million&40&37 (0)&39 (14)\\\hline
Kiddopedia&339 Million&717 Thousand&116&53 (0)&65 (2)\\\hline
Baby Einstein&575 Million&768 Thousand&124&23 (0)&37 (9)\\\hline
Fun Kids Planet&828 Million&1 Million&121&86 (3)&107 (26)\\\hline
ChuChuTV Surprise Eggs Learning Videos&4 Billion&7 Million&64& 16 (0)&23 (5)\\\hline
ChuChu TV Nursery Rhymes \& Kids Songs&36 Billion&54 Million&59&21 (1)&30 (13)\\\hline
Funny Kids Playtime with Jade \& James - ToysReview&151 Thousand&1 Thousand&29&24 (3)&27 (9)\\\hline
Dipo Dipo&8 Million&19 Thousand&25&0 (0)& 1 (1)\\\hline
Two Kids TV&154 Million&308 Thousand&11&1 (0)&1 (0)\\\hline
Cupcake Squad&788 Million&2 Million&1&1 (0)& 0 (0)\\\hline

 \end{tabular}
 \vspace{0.5cm}
 \caption{List of YouTube channels considered. Channel statistics reflect data as on  20 Jan 2022. Numbers in braces () indicate highly inappropriate taboo words' count.}
 \label{tab:channels}
\end{table}
\end{center}

\section{Data and Design Considerations}

\subsection{Data Set}
We create a new data set by collecting videos from top YouTube Kids (YTK) channels. In order to construct a data set highly relevant for kids, among the vast number of channels on YTK, we focus on the top-ranked channels based on their popularity (i.e., number of views). We use two rankings from Wall Street Journal \footnote{https://www.wsj.com/articles/kids-love-these-youtube-channels-who-creates-them-is-a-mystery-11554975000} and Statista  \footnote{https://www.statista.com/statistics/785626/most-popular-youtube-children-channels-ranked-by-subscribers/} to get a few very popular channels. 

For these channels, we next retrieve all English language videos in those channels using a widely used library\footnote{https://github.com/ytdl-org/youtube-dl}. Our data set comprises a total of 7,013 videos extracted across 24 channels that fall under the YouTube Kids category. We consider both the transcriptions provided by the YouTube API and Amazon Transcribe in our experiments. Further details regarding the data set are listed in Table \ref{tab:channels}.


Some of the videos involved music and rhymes, some involved interaction between two or more characters, and some involved no form of speech at all. Despite there being no verbal interactions in some of these videos, we found that in many of these cases, the services nevertheless produced transcriptions, which in some cases contained toxic language and hate speech, in direct violation of YouTube Kids' guidelines. \footnote{https://support.google.com/youtube/answer/9528076?hl=en}

     
   


\subsection{Speech to Text Methods}
We next give a brief description of the two speech-to-text transcript services that we consider. \begin{compactitem}

    \item Amazon (AWS) Transcription:
    As per the description on Amazon website\footnote{https://aws.amazon.com/transcribe/}, AWS Transcribe uses a deep learning method to convert speech to text. In addition to subtitling, Amazon Transcribe also generates metadata including the number of speakers and which transcription was a result of the speech by which speaker. For generating AWS transcriptions, we first obtained the audio of the YouTube videos. Then we used AWS transcribe service to transcribe the audio files to text.

    \item YouTube (Google) Transcription:
    In addition to AWS, we also consider YouTube transcriptions. YouTube transcriptions are created when a video is uploaded. As per YouTube\footnote{https://support.google.com/youtube/answer/6373554}, automatic captions are generated by machine learning algorithms, so the quality of captions may vary across videos. Though there are a number of languages for which captions could be available, for the analysis in this paper, we only consider videos for which English language transcripts (also called captions) are available via YouTube API\footnote{https://developers.google.com/youtube/v3/docs/captions}.

    
\end{compactitem}

\subsection{Transcription Quality}

It may well be the case that the overall transcription quality of our data set is low and the presence of taboo-words are mere artifacts of noisy outputs. We first thus validate the quality of transcripts. 

\subsubsection{Mutual Agreement}

Let $\mathcal{T}_i^{\mathcal{A}_j}$ denote the transcript obtained from video $v_i$ using transcription algorithm $\mathcal{A}_j$. Let $\mathcal{N}(w_k, \mathcal{T}_i)$ denote the total number of occurrences of the word $w_k$ in transcript $\mathcal{T}_i$. Further, let $\mathcal{V} (\mathcal{T}_i)$ denote the vocabulary of a transcript $\mathcal{T}_i$. The mutual agreement between two transcripts of the same video is the fraction of words that have appeared identical number of times over both transcripts:\\
\emph{MA} ($\mathcal{T}_i^{\mathcal{A}_j}$, $\mathcal{T}_i^{\mathcal{A}_{j'}}$) = $\frac{\Sigma_{w_k \in \mathcal{V} } \mathbbm{I} (\mathcal{N}(w_k, \mathcal{T}_i^{\mathcal{A}_{j}}) = \mathcal{N}(w_k, \mathcal{T}_i^{\mathcal{A}_{j'}}) )}{|\mathcal{V}|}$
where $\mathcal{V} = \mathcal{V}(\mathcal{T}_i^{\mathcal{A}_j}) \cup \mathcal{V}(\mathcal{T}_i^{\mathcal{A}_{j'}})$, i.e., the union of the vocabularies of the transcripts generated by algorithms $\mathcal{A}_{j}$ and $\mathcal{A}_{j'}$ when applied to video $v_i$.

\begin{figure}[htb]
\centering
\includegraphics[trim={0 0 0 0},clip, height=1.9in]{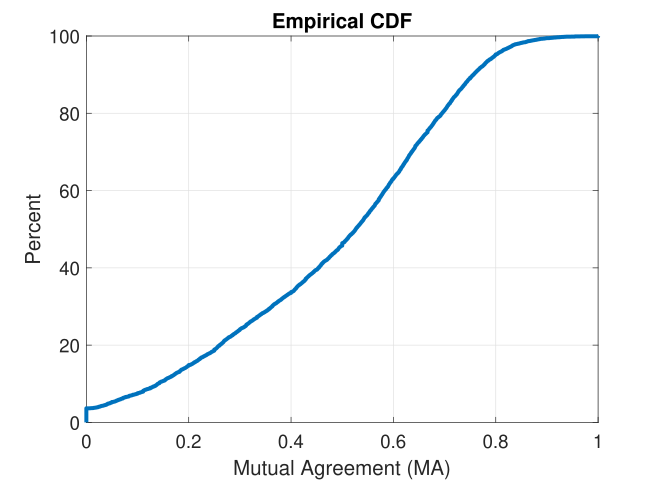}
\caption{A CDF plot of mutual agreement in our data set between Amazon transcribe and YouTube speech-to-text.} 
\label{fig:MutualAgreement}
\end{figure}

Our intuition is if both transcription algorithms perform a reliable job on a given video, the mutual agreement value will be high. We note that this is a quite stringent criterion as every single error while transcribing a video may contribute to a reduced \emph{MA}. Figure~\ref{fig:MutualAgreement} indicates that more than 42\% of the videos have mutual agreement higher than 50\%. In addition, we conduct a human inspection of the videos and confirm that several videos in our data set have overall high-quality transcripts.

\subsection{Developing a Set of Taboo-words for Kids}

Deciding on the set of taboo-words is one of the major design considerations in this project. Several factors such as subjectivity, cultural contexts, and audience can determine if a word is perceived as inappropriate or not. Consequently, there exists no broad consensus on hate lexicons. As a starting point, we select a well-known, publicly available hate lexicon\footnote{Available at \url{https://www.cs.cmu.edu/~biglou/resources/bad-words.txt}} containing more than 1,300 words (denoted as $\mathcal{H}_1$). While describing this lexicon as a reasonable starting point to block or filter offensive content, curators of this lexicon acknowledge the subjectivity of this lexicon stating that the list contains words that many people may \textbf{not} find offensive. 

Our second choice of the lexicon is strongly grounded in prior literature in developmental psychology~\cite{sutton1978psychosexual,jay2013child}. We obtain a list of 76 words presented in~\cite{jay1992cursing} as taboo-words for children. These words are collected from actual usage of these words by children within the age range of 1--10 in a field study~\cite{jay1992cursing} (denoted as $\mathcal{H}_2$).

We note that both $\mathcal{H}_1$ and $\mathcal{H}_2$ complement each other and combining them may have certain merits. Since $\mathcal{H}_2$ consists of words actually used by kids in a field study, it precludes certain inappropriate words with strong sexual connotations as these words require adult-level understanding. For instance, unlike $\mathcal{H}_1$, words like \texttt{cocksucker} or \texttt{rape} are absent in $\mathcal{H}_2$. In our work, we are focusing on content hallucinations from ASR systems. Thus casting a net wider than $\mathcal{H}_2$ has understandable benefits (in fact, as shown in Table~\ref{tab:example}, one of the content hallucination indeed produces the taboo-word \texttt{rape}). 

While words belonging to $\mathcal{H}_2$ serve an important purpose to give us a broad understanding of what could be construed as inappropriate for kids, we note that some of these words could be heavily context-dependent. For instance, the word \texttt{dog} may be used in a completely non-taboo scenario. Similarly, we find that  $\mathcal{H}_1$ also contains certain words that can be used in non-taboo scenarios. For example, the word \texttt{killer} can be used in a completely harmless context of \texttt{killer whale}.

\begin{figure*}[htb!]
    \centering
    \subfloat[\centering Amazon transcribe]{{\includegraphics[width=0.40\textwidth]{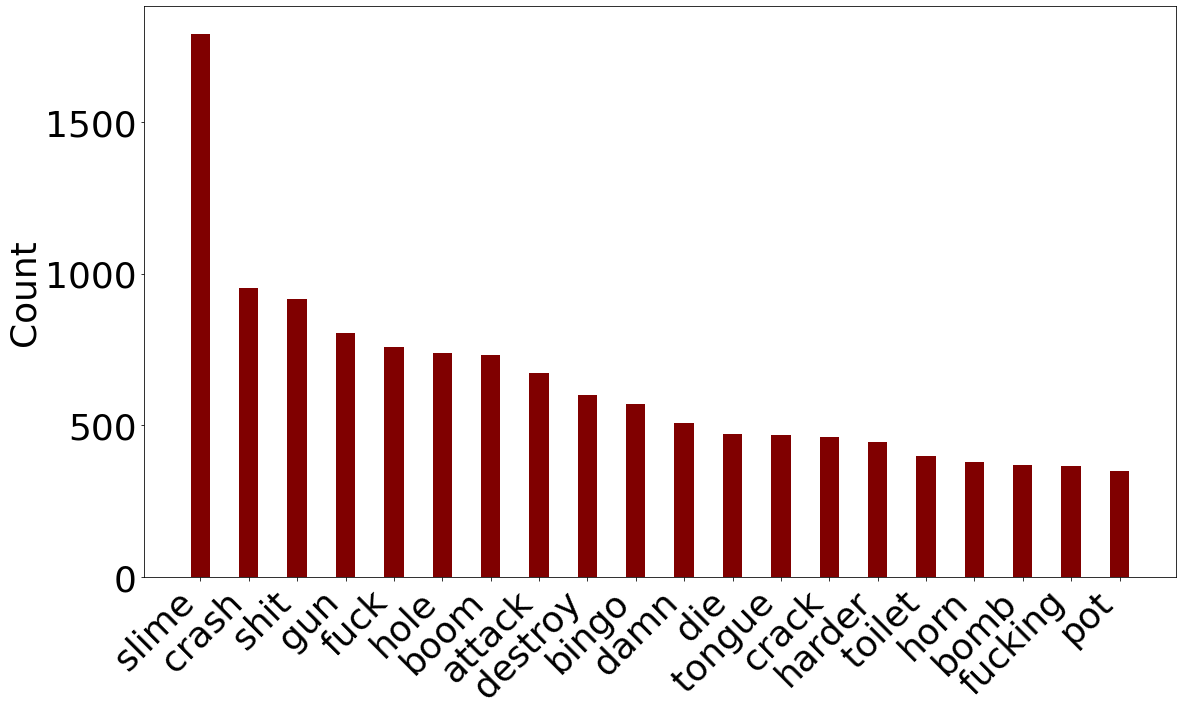} }}%
    \qquad
    \subfloat[\centering Google speech-to-text]{{\includegraphics[width=0.40\textwidth]{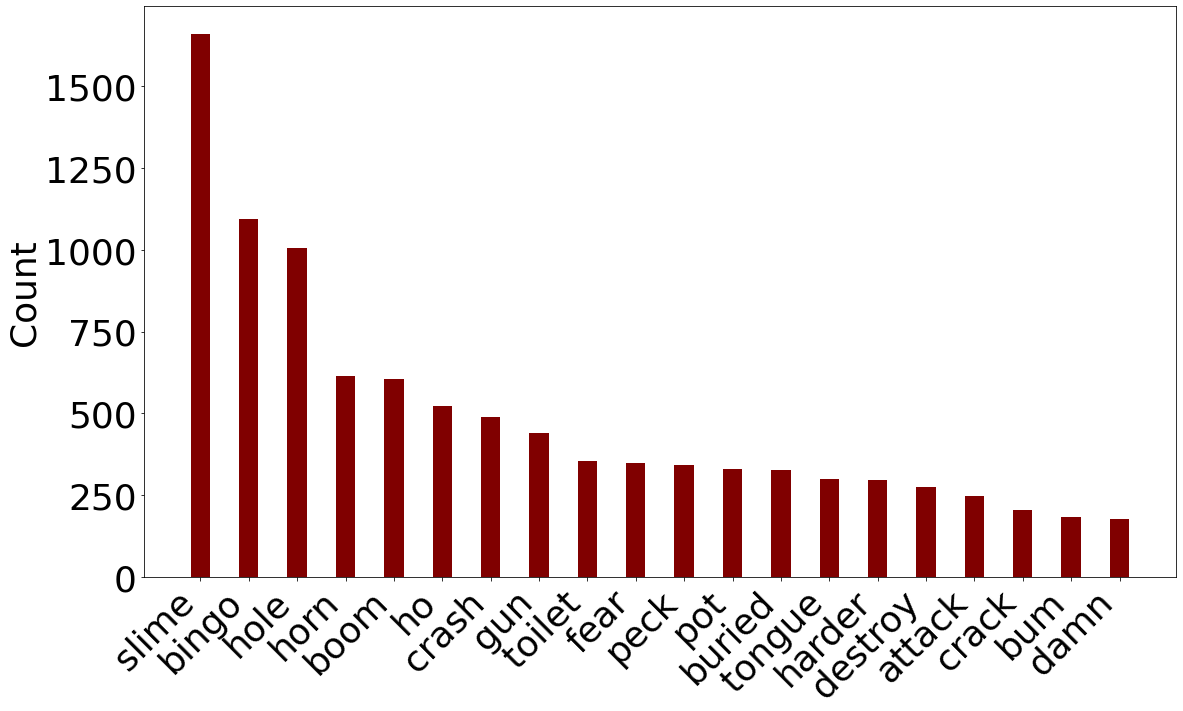} }}%
    \caption{Top twenty taboo-words (potentially) hallucinated by Amazon transcribe and Google speech-to-text in our data set.}%
    \label{fig:SatSolver}%
\end{figure*}

\begin{figure*}[htb!]
    \centering
    \subfloat[\centering Amazon transcribe]{{\includegraphics[width=0.40\textwidth]{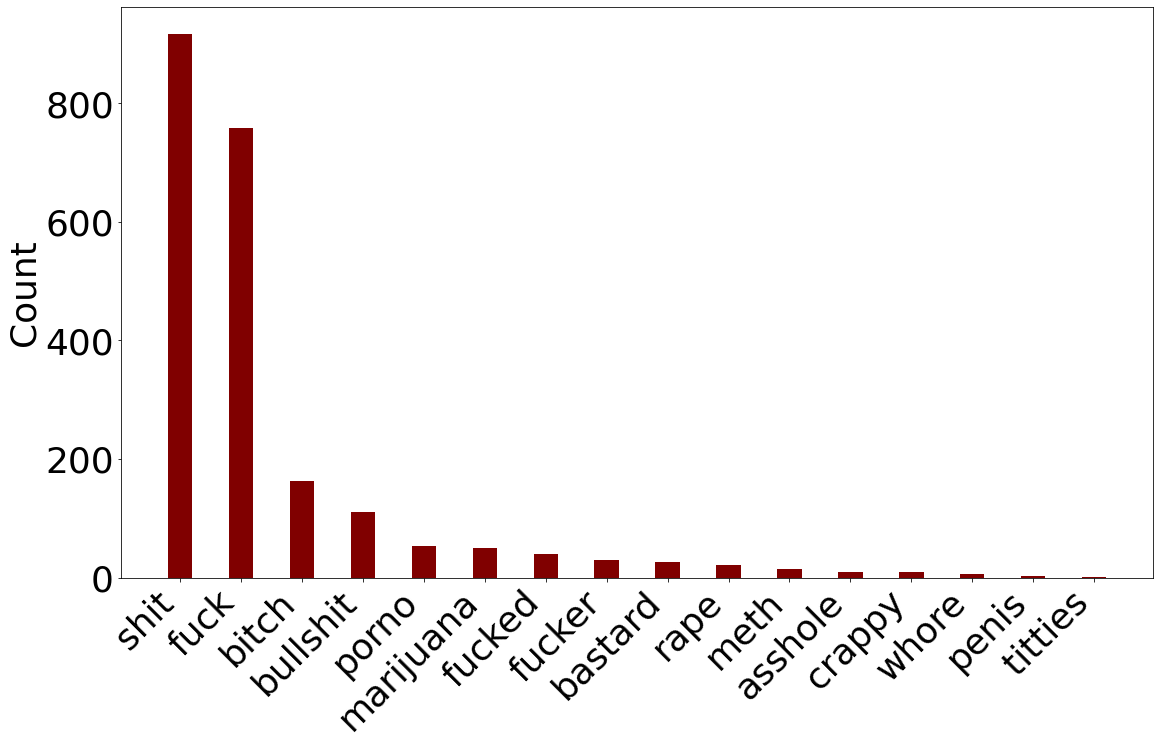} }}%
    \qquad
    \subfloat[\centering Google speech-to-text]{{\includegraphics[width=0.40\textwidth]{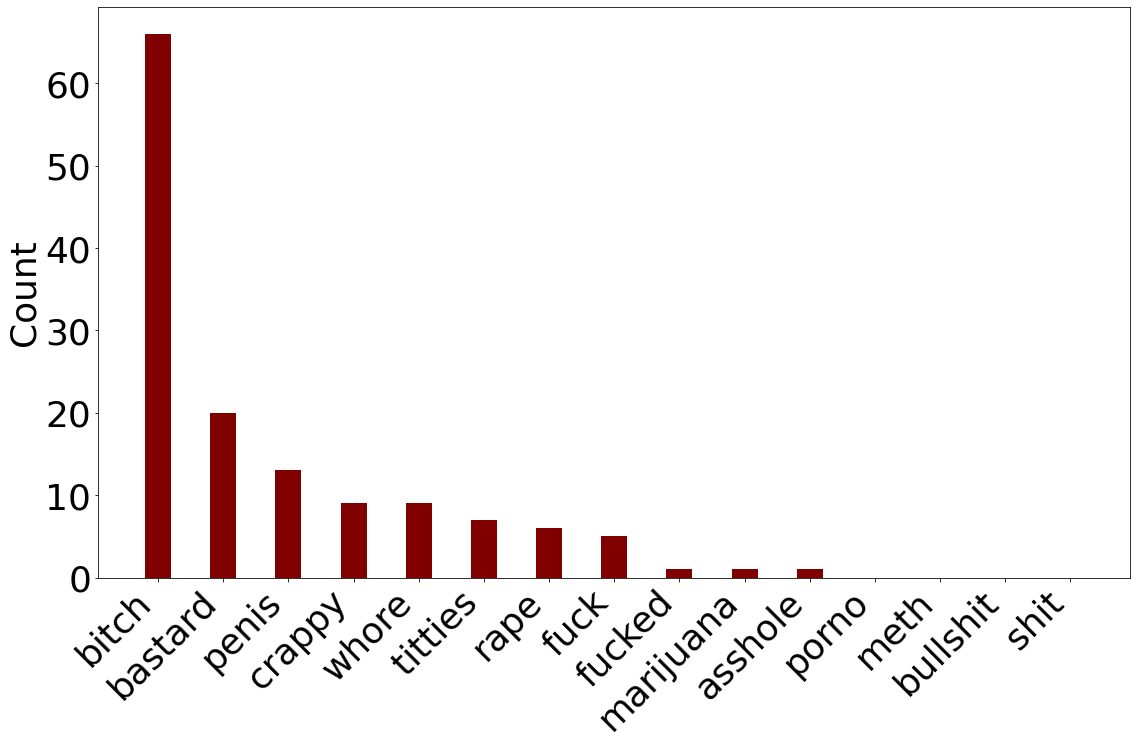} }}%
    \caption{Top few highly inappropriate taboo-words (potentially) hallucinated by Amazon transcribe and Google speech-to-text in our data set.}%
    \label{fig:highly_inappropriate_taboor_wors}%
\end{figure*}

We thus combine both $\mathcal{H}_1$ and $\mathcal{H}_2$ and analyze to what extent these words are present in children's movies. We construct a data set, $\mathcal{D}_{\mathit{Disney}\emph{-}\mathit{Pixar}}$, consisting of English subtitles\footnote{Subtitles are obtained from \url{Subscene.org}} of all movies released in or after 2000 with an MPAA movie rating of (G) implying that these movies are certified as safe for the general audience and nothing would offend parents for viewing by children. Overall, we obtained 57 movies. Our choice of these well-regarded movie franchises with MPAA certifications of (G) serves two key purposes. First, a wide variety of entertainment content targeted for kids ensures that the corpus consists of a rich set of kids' entertainment contexts. Second, the MPAA rating indicates that these contexts are certified as appropriate for children. Our intuition is if a word $w \in \mathcal{H}_1 \cup \mathcal{H}_2$ appears on multiple occasions in $\mathcal{D}_{\mathit{Disney}\emph{-}\mathit{Pixar}}$, it possibly indicates that that $w$ can be used in non-taboo scenarios for kids. Table~\ref{tab:movieLexicon} shows top 10 words from $\mathcal{H}_1$ and $\mathcal{H}_2$ that have appeared at least five or more times in $\mathcal{D}_{\mathit{Disney}\emph{-}\mathit{Pixar}}$. We further note that in $\mathcal{D}_{\mathit{Disney}\emph{-}\mathit{Pixar}}$, we observe zero mentions of words with strong sexual connotation such as \texttt{rape} and \texttt{fuck} and scatological references such as \texttt{shit}.  

\begin{table}[htb]

{
\scriptsize
\begin{center}
     \begin{tabular}{| p{3.5cm}  | p{3,5cm} |}
    \hline
    \Tstrut $\mathcal{H}_1$  & $\mathcal{H}_2$ \\
     \hline \Tstrut kid,
fairy,
fairies,
girls,
dead,
fight,
god,
bigger,
stupid,
shoot \Bstrut&      dog,
god,
stupid,
hell,
pig,
fat,
chicken,
nuts,
silly,
butt \\
    \hline
    \end{tabular}
\end{center}
\caption{{Top ten words (ranked by frequency) from $\mathcal{H}_1$ and $\mathcal{H}_2$} present in $\mathcal{D}_{\mathit{Disney}\emph{-}\mathit{Pixar}}$. $\mathcal{D}_{\mathit{Disney}\emph{-}\mathit{Pixar}}$ consists of English subtitles of all Disney and Pixar movies with MPAA rating (G) released in or after 2000.}
\label{tab:movieLexicon}}
\end{table}

We remove all words from $\mathcal{H}_1 \cup \mathcal{H}_2$ that have appeared for five or more times in $\mathcal{D}_{\mathit{Disney}\emph{-}\mathit{Pixar}}$. In addition, we manually remove words indicating nationality of a person (e.g., \texttt{Italian}, \texttt{American}) or religious words (e.g., \texttt{Muslim}). After this step, our combined lexicon of taboo-words, $\mathcal{H}$, consists of 1,301 words.   


\subsection{Extent of Presence of Taboo-words}

Overall, we find that 330 taboo-words were present in YouTube transcripts and 386 taboo-words were present in AWS transcripts in our data set. Figure~\ref{fig:SatSolver} presents the distribution of the top twenty words belonging to $\mathcal{H}$ present in our transcript data set. We observe the considerable presence of inappropriate taboo-words such as \texttt{shit} in transcripts generated by Amazon transcribe and Google speech-to-text. 

\begin{figure*}[htb]
    \centering
    \subfloat[\centering {\bf \href{https://www.youtube.com/watch?v=8aH0ttHyT-M&t=137.92s}{\color{blue}SimpleKidsCrafts}}: $\text{venus} \longrightarrow penis$]{{\includegraphics[width=0.40\textwidth]{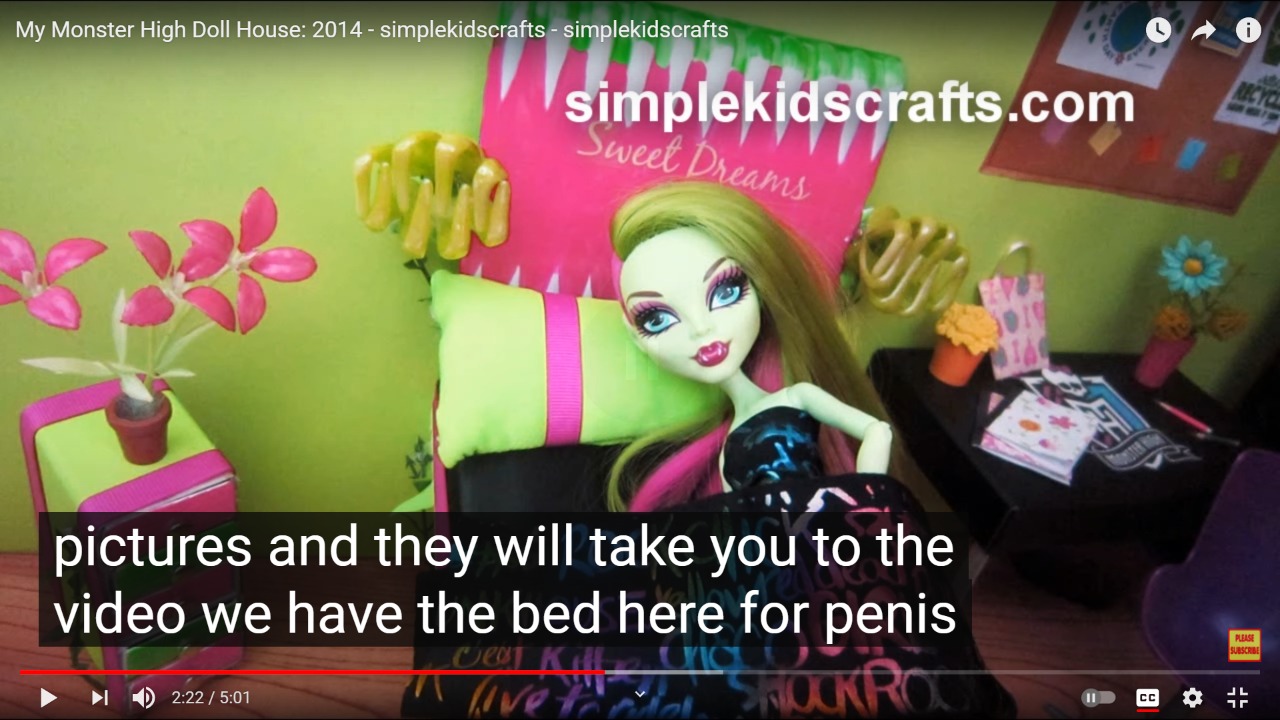} }}%
    \qquad
    \subfloat[\centering {\bf \href{https://www.youtube.com/watch?v=vtI1Jq4LHRQ&t=492.79s}{\color{blue}Ryan's World}}: $\text{corn}  \longrightarrow porn$]{{\includegraphics[width=0.40\textwidth]{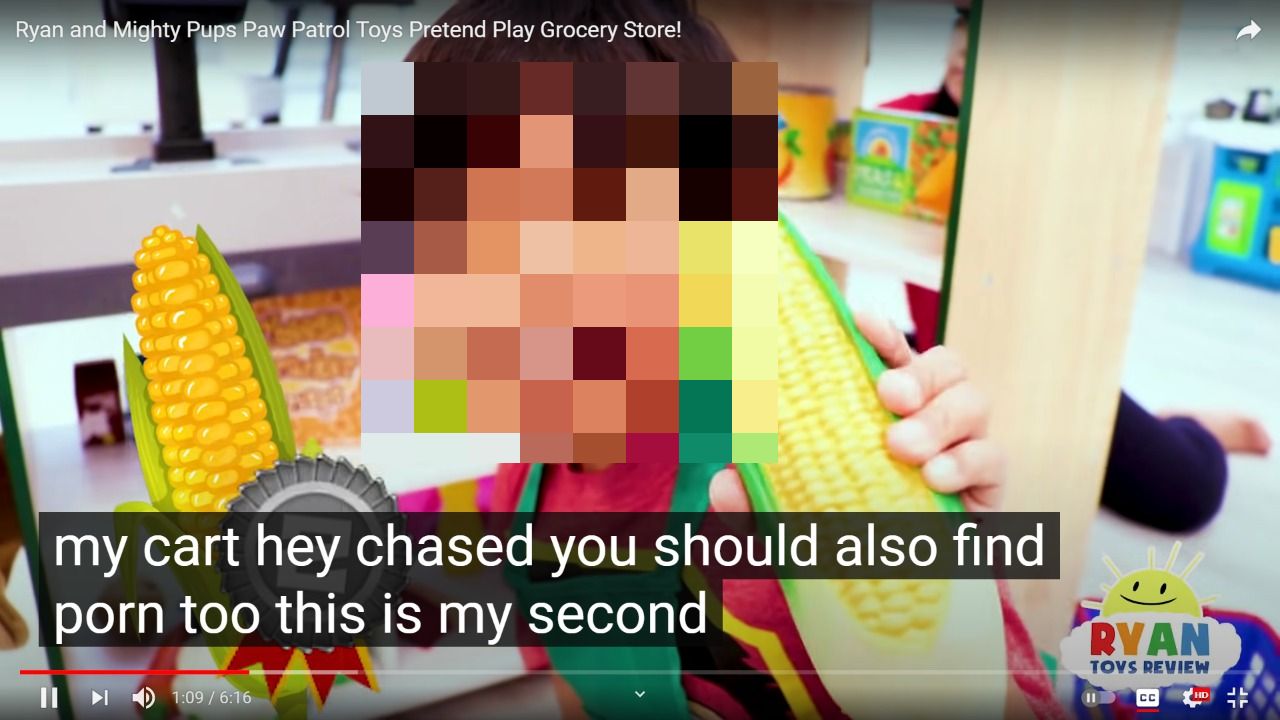} }}%
    
    \subfloat[\centering {\bf \href{https://www.youtube.com/watch?v=MjxPIbLtmHg&t=67.17s}{\color{blue}Ryan's World}}: $\text{that is} \longrightarrow panties$]{{\includegraphics[width=0.40\textwidth]{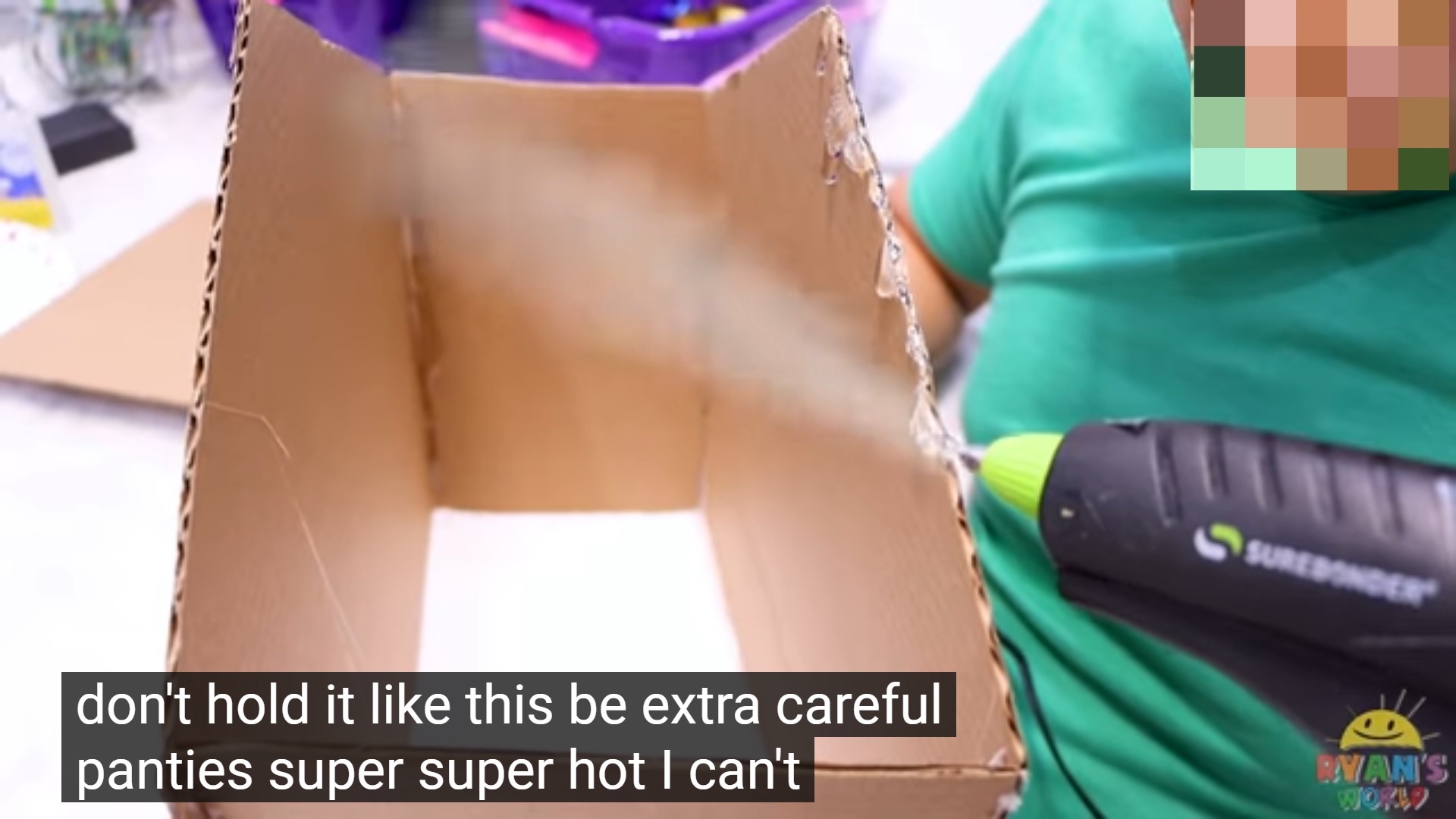}}}
    \qquad
    \subfloat[\centering {\bf \href{https://www.youtube.com/watch?v=L2VLJBbFaBM&t=1369.94s}{\color{blue}TRT TV}}: $ \text{buster} \longrightarrow bastard$]{{\includegraphics[width=0.40\textwidth]{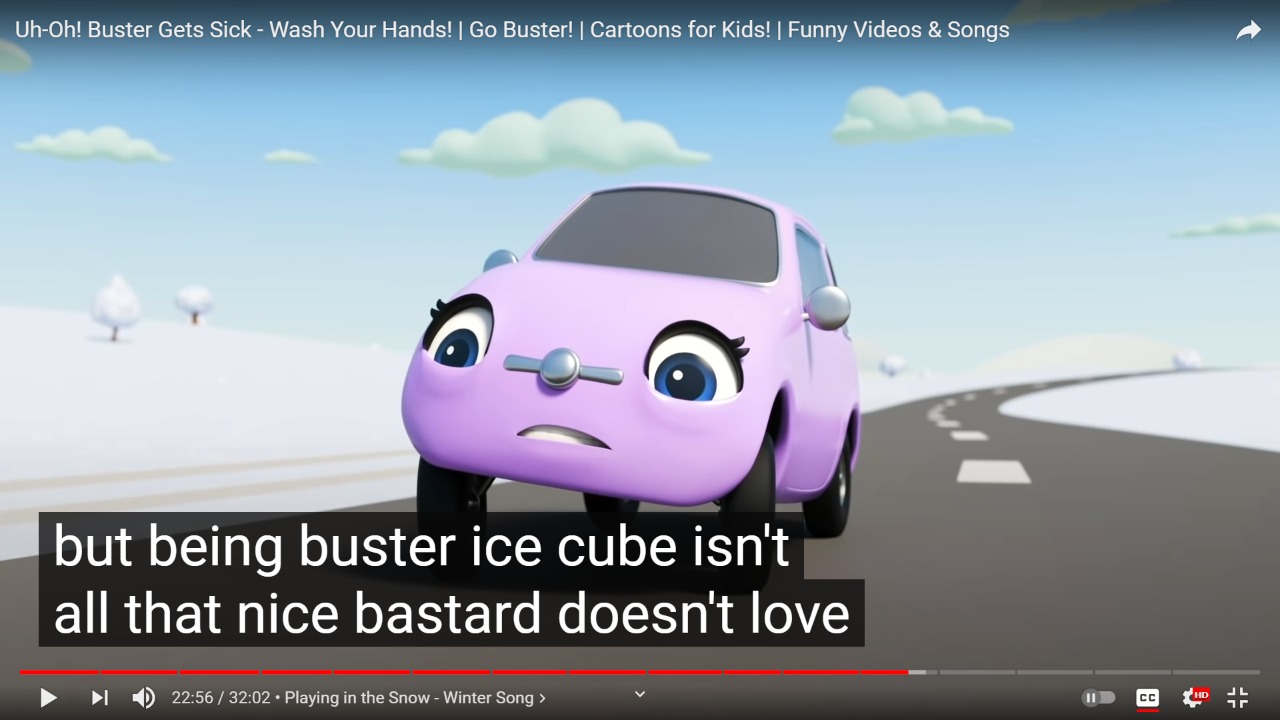} }}%
    
    \subfloat[\centering {\bf \href{https://www.youtube.com/watch?v=zEGTw8UBWfE&t=1944s}{\color{blue}Rob The Robot - Learning Videos For Children}}: $ \text{brave} \longrightarrow rape$]{{\includegraphics[width=0.40\textwidth]{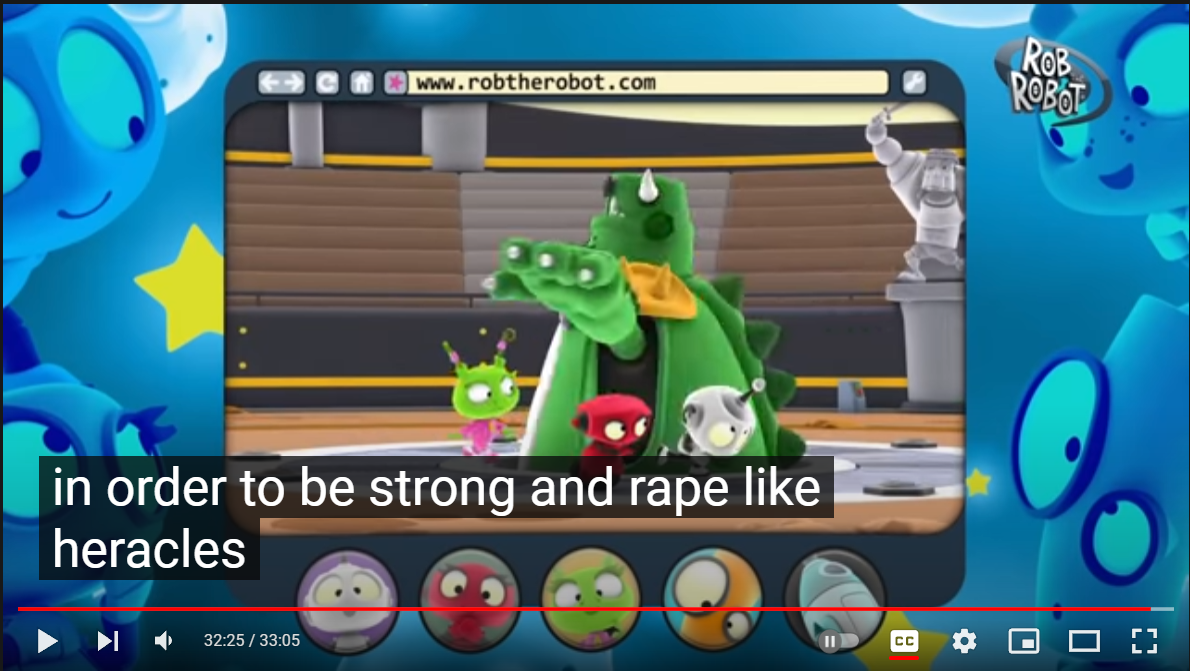} }}%
    \qquad
    \subfloat[\centering {\bf \href{https://www.youtube.com/watch?v=N12V9CJIoD4&t=10.9s}{\color{blue}Ryan's World}}: $ \text{combo} \longrightarrow condom$ ]{{\includegraphics[width=0.40\textwidth]{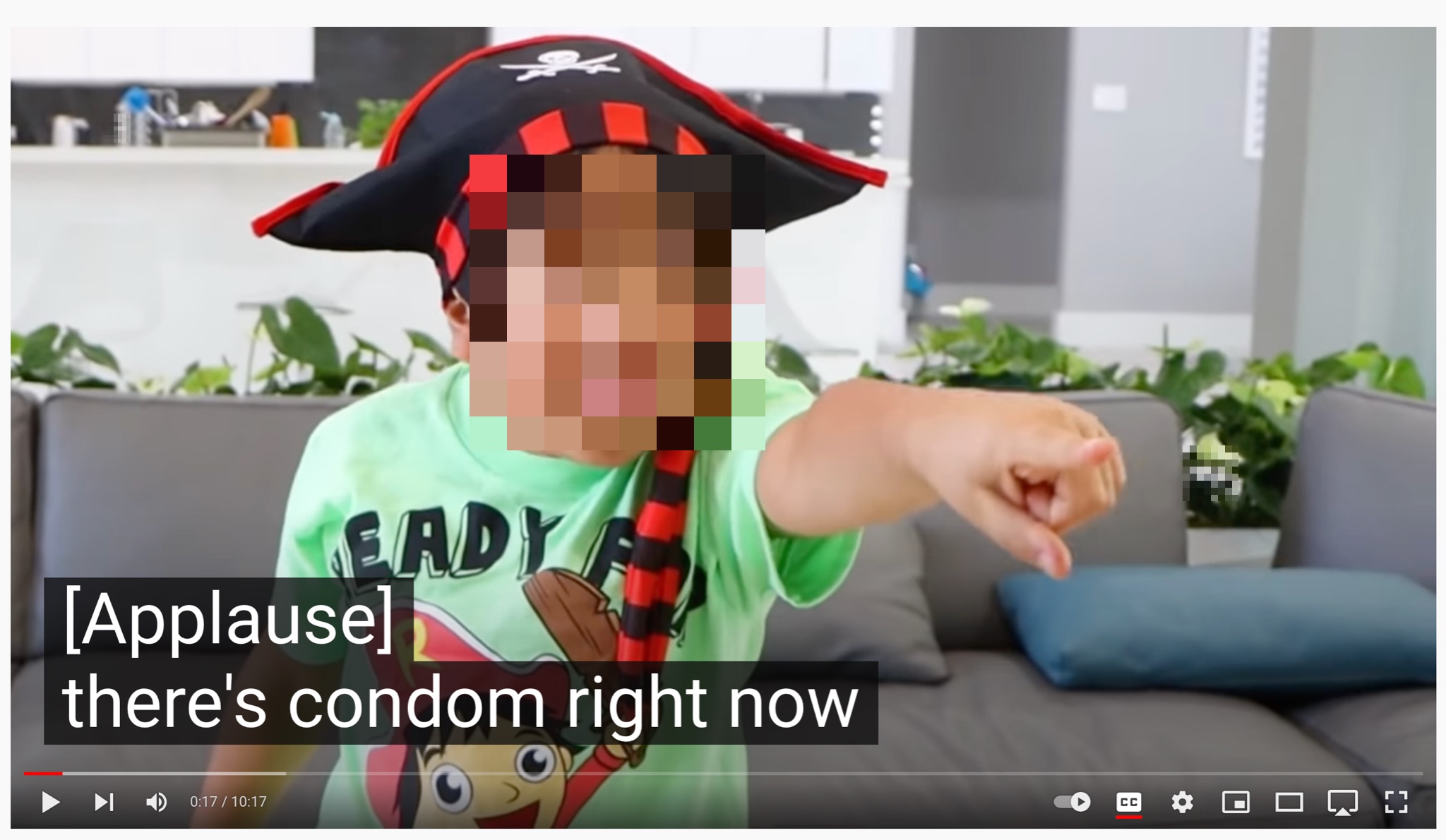} }}%
    
    \caption{Examples of hallucinated taboo-words from YouTube along with corresponding ground truths.}%
    \label{fig:example}%
\end{figure*}

While Figure~\ref{fig:SatSolver} indicates a worrisome finding of considerable presence of taboo-words in video transcripts, some of these words may not be hallucinated and could be present in contexts safe for kids. For example, \texttt{ho}, which also implies the disparaging and offensive slang \texttt{whore}, could be simply present in a Santa Claus video. 
We thus manually inspected $\mathcal{H}$ and shortlisted a set of 16 words that are unambiguously inappropriate and analyzed their presence in the video transcripts. As shown in Figure~\ref{fig:highly_inappropriate_taboor_wors}, we find that the video transcripts also exhibit substantial presence of these highly inappropriate words. In fact, Table~\ref{tab:channels} indicates that nearly one in ten videos contains at least one or more of these highly inappropriate taboo-words in transcription generated either by AWS or Google speech-to-text. In addition, we show some less frequent but highly inappropriate words in Figure \ref{fig:example}.


\textbf{\emph{RQ 1:}} \emph{Are these taboo-words indeed hallucinated, or are they indeed present in the actual audio content?} 

We first randomly sample 100 contexts containing five highly inappropriate taboo-words (\{\texttt{shit}, \texttt{fuck}, \texttt{crap}, \texttt{rape}, \texttt{ass}\}) and manually inspect if these words are hallucinated or if they are indeed present in the source. Two annotators independently listened to the audio clips and confirmed that none of the taboo-words occurred in the actual audio. Upon manual inspection, we identify the following high-level potential factors to these content hallucinations: (1) background music; (2) baby talk; (3) kids' speech; (4) ESL (English as Second Language speakers) speakers; and (5) songs and rhymes. Note that, we do not intend to be formal or exhaustive, but rather to be illustrative of the broad range of potential reasons that can cause such hallucination.

\begin{figure}[htb]
\centering
\includegraphics[trim={0 0 0 0},clip, height=2.1in]{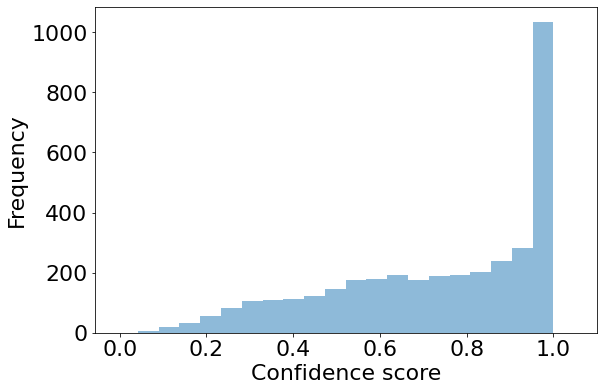}
\caption{Plot of the confidence scores of all taboo-words and their frequency from Amazon transcribe.} 
\label{fig:confidence_inappropriate}
\end{figure}

\textbf{\emph{RQ 2:}} \emph{How confident is the transcription method while hallucinating a taboo-word?} 
Amazon transcribe presents a confidence estimate of individual words. Figure~\ref{fig:confidence_inappropriate} indicates that a vast number of taboo-words were high-confidence predictions of this system.  

Once we establish that these taboo-words are largely absent in the audio inputs and the transcription methods often produce them while reliably transcribing a large part of the adjacent audio inputs, we turn our focus into creating a data set consisting of challenging audio inputs where high-performance commercial ASR systems hallucinate taboo-words.

We construct a data set, $\mathcal{D}_\mathit{taboo}$, consisting of 284 YouTube transcriptions and 368 Amazon Transcribe transcriptions that satisfy the following conditions: (1) the snippet contains a taboo-word; and (2) both transcription algorithms exhibit considerable agreement within transcribing the snippet (algorithm sketch is presented in the project page). All samples contain consensus labels from two annotators.  

\section{Corrective Method} \label{Corrective Method}

Our method to correct taboo-words from video transcripts  relies on cloze tasks performed by recent high-performance language models such as \texttt{BERT}~\cite{devlin2018bert}, XLM~\cite{chi2021xlme}, DistilBERT~\cite{DBLP:journals/corr/abs-1910-01108}, XLNet~\cite{DBLP:journals/corr/abs-1906-08237} and Megatron~\cite{DBLP:journals/corr/abs-1909-08053} . When presented with a sentence (or a sentence stem) with a missing word, a cloze task is essentially a fill-in-the-blank task. For instance, in the following cloze task: \emph{During the} \texttt{[MASK]}\emph{, it rains a lot}, \texttt{monsoon} is a likely completion for the missing word. \texttt{BERT}'s masked word prediction has a direct parallel to cloze task introduced in the psycholinguistics literature~\cite{taylor1953cloze}. \texttt{BERT}'s cloze task has been previously used in the (1) extracting relational knowledge~\cite{petroni-etal-2019-language}; (2) mining political insights~\cite{electionKhudaBukhsh}; (3) assessing the quality of translation~\cite{bert-score}; and (4) estimating linguistic quality~\cite{sarkar-etal-2020-non}. 

Our method's intuitions are the following. It is highly likely that the ground truth is phonetically (and lexically) similar to the taboo-word (e.g., $\langle\texttt{crap}, \texttt{crab}\rangle$; $\langle\texttt{seat}, \texttt{shit}\rangle$; $\langle\texttt{rape}, \texttt{rake}\rangle$). Our method thus first constructs a set of candidate words based on some notion of proximity (lexical or phonetic). Next, it conducts a constrained cloze task using a language model that only considers the candidate set as potential completions. Assuming that the transcriptions for the context minus the taboo-word are reliable, we expect that the context would be able to guide the language models towards the correct completion. 

Let \texttt{LM}$_\mathit{cloze} (w, \mathcal{S})$ denote the completion probability of the word $w$ when a language model, \texttt{LM}, has a masked cloze task $\mathcal{S}$ as input. When we have a text snippet with a taboo-word, we construct $S$ by masking the taboo-word. For example, if our snippet is the following: \emph{I love to eat \textbf{\textcolor{red}{crap}} and lobster for dinner.}, our cloze task will be \emph{I love to eat [\texttt{MASK}] and lobster for dinner.}  Let our candidate set, $\mathcal{C} = \{\texttt{crab}, \texttt{crap}, \texttt{craft}\}$, be consisting of words that are similar to the taboo-word based on some similarity measure.


Our corrective method will output $c^{*}$ obtained by:

\begin{equation*}
    c^{*} = \argmax_{c~\in~\mathcal{C}} \texttt{LM}_{\mathit{cloze}}(c, \mathcal{S})
\end{equation*}

For language models, we consider several well-known high-performance language models: \texttt{BERT}, \texttt{XLM}, \texttt{XLNet}, \texttt{DistilBERT}, and \texttt{Megatron}. For a given taboo-word, we generate the candidate set using two different approaches. In our first approach, we consider words with a low Levenshtein distance with the taboo-word. In our second approach, we consider highly phonetically similar words to the taboo-word. In order to conduct a fair assessment of the language models, we restrict our experiments to samples where (1) single-word substitutions would suffice to fix the error; and (2) the ground truth is included in the language model's vocabulary.

Table~\ref{tab:ClozeResults} summarizes our correction results. \textbf{P@1} performance indicates the fraction of instances where the top predicted completion in the cloze test is indeed the ground truth. Following, standard practice~\cite{petroni-etal-2019-language}, we also report the \textbf{P@5} and \textbf{P@10} performance where \textbf{P@K} performance indicates that the top $K$ completions for the cloze test contains the ground truth.  The computing infrastructure used for running experiments is described in the project page along with information on the hyper-parameters used for each model.

We find that among all the combinations, $\langle \texttt{Megatron}, \mathit{Levenshtein} \rangle$ 
obtained the best performance on the data, fixing more than 25\% of the errors produced by Amazon Transcribe and over 28\% of the errors hallucinated by Google speech-to-text. We were not surprised by our modest success at fixing these hallucinated taboo-words. While the inner workings of Amazon transcribe and Google speech-to-text are opaque, there exists substantial literature on ASR systems leveraging language models. Furthermore, here, we are attempting to fix non-trivial, extremely challenging transcription errors hallucinated by industry-scale, commercial solutions. We are rather encouraged by a high \textbf{P@5} performance indicating that a human-in-the-loop setting can be aided by our method for corrective purposes.   

We next turn our focus to some of the correctly fixed examples. As shown in Table~\ref{tab:Fixed}, we notice that when sentences are well-formed and present with enough context, language models are often successful at fixing the error. However, given the nature of kids' channels, many of the transcripts contain incoherent, ill-formed sentences, thus making it extremely challenging for LMs to correctly predict the masked word. 

\renewcommand{\tabcolsep}{1mm}
\begin{table}[hbt]
{
\small
\begin{center}
     \begin{tabular}{|p{0.34\textwidth}|p{0.56\textwidth}|}
    \hline
\textcolor{red}{rape} $|$ \textcolor{blue}{brave} &     monsters in order to be strong and \textbf{\textcolor{red}{rape}} like heracles we even had a chariot race in order to be fast like heracles but when orbit rescued emma we realized theres more to a hero than just being\\
    \hline
\textcolor{red}{bitch} $|$ \textcolor{blue}{beach} &     glasses so you can see that and then here we have his sandals his sandals are completely made out of plastic they are orange and they have the same flames at the top and then we have a little \textbf{\textcolor{red}{bitch}} towel that came with him and what is really cool about this towel is a motif which\\
    \hline
\textcolor{red}{crap} $|$ \textcolor{blue}{craft} &     if you have any requests or \textbf{\textcolor{red}{crap}} ideas that you would like us to explore kindly send us an email\\
    \hline
    \end{tabular}
    
\end{center}
\caption{{Examples of correctly fixed snippets. Hallucinated taboo-word is marked with red, ground truth is marked with blue.}}
\label{tab:Fixed}}
\vspace{-0.3cm}
\end{table}

\renewcommand{\tabcolsep}{1mm}
\begin{table}[hbt]
{
\small
\begin{center}
     \begin{tabular}{|p{0.34\textwidth}|p{0.56\textwidth}|}
    \hline
\textcolor{red}{cocktail} $|$ \textcolor{blue}{copter} $|$ \textcolor{gray}{social} &     also exceptionally strong for their size they can lift 10 to 50 times their own weight thats like being a little hero that can lift their own \textbf{\textcolor{red}{cocktail}} over their head other cool features\\
    \hline
\textcolor{red}{penis} $|$ \textcolor{blue}{venus}  $|$ \textcolor{gray}{pets} &     you need is in the description but as were passing through the pictures you can click on the pictures and they will take you to the video we have the bed here for \textbf{\textcolor{red}{penis}} and the side drawers as well and here we have the arts and crafts studio which is basically everything from\\
    \hline
\textcolor{red}{bastard} $|$ \textcolor{blue}{buster} $|$ \textcolor{gray}{stars} &   indeed if you are in trouble then who will help you out here at super \textbf{\textcolor{red}{bastard}} quest without a doubt\\
    \hline
    \end{tabular}
    
\end{center}
\caption{{Examples of incorrectly fixed snippets. Hallucinated taboo-word is marked with red, ground truth is marked with blue, top prediction using cloze test is marked with gray.}}
\label{tab:NotFixed}}
\vspace{-0.3cm}
\end{table}










\begin{table}[ht!]
\small
\centering
\begin{tabular}{|l|c|l|l|l|}
\hline
Corrective method                                                                               & Transcription algorithm & P@1     & \multicolumn{1}{c|}{P@5}     & \multicolumn{1}{c|}{P@10}            \\ \hline
$\langle\texttt{BERT}, Levenshtein\rangle$                                                      & Amazon                  & 19.14\% & {40.67\%} & {45.93\%}   \\ \hline
$\langle\texttt{BERT}, Phonetic\rangle$                                                         & Amazon                  & 7.66\%   & {25.84\%} & {35.41\%}   \\ \hline
$\langle\texttt{XLM}, Levenshtein\rangle$                                                       & Amazon                  & 6.22\% & {24.41\%} & {40.67\%}   \\ \hline
$\langle\texttt{XLM}, Phonetic\rangle$                                                          & Amazon                  & 6.22\%  & {22.01\%} & {35.89\%}  \\ \hline
$\langle\texttt{XLNet}, Levenshtein\rangle$      & Amazon                  & 5.26\%  & 22.97\%                      & 39.71\%                       \\ \hline
$\langle\texttt{XLNet}, Phonetic\rangle$                                                        & Amazon                  & 4.79\%  & 15.79\%                      & 33.97\%                      \\ \hline
$\langle\texttt{DistilBERT}, Levenshtein\rangle$ & Amazon                  & 16.75\% & 38.76\%                      & 44.98\%   \\ \hline
$\langle\texttt{DistilBERT}, Phonetic\rangle$                                                   & Amazon                  & 8.13\%  & 25.36\%                      & 35.41\%                  \\ \hline
$\cellcolor{blue!15}\langle\texttt{Megatron}, Levenshtein\rangle$   & Amazon                  & \cellcolor{blue!15}25.36\% & 40.67\%                      & 44.98\%  \\ \hline
$\langle\texttt{Megatron}, Phonetic\rangle$                                                     & Amazon                  & 10.53\% & 27.75\%                      & 34.93\%                      \\ \hline
$\langle\texttt{BERT}, Levenshtein\rangle$                                                      & YouTube                 & 22.35\% & \multicolumn{1}{c|}{37.65\%} & \multicolumn{1}{c|}{44.12\%}  \\ \hline
$\langle\texttt{BERT}, Phonetic\rangle$                                                         & YouTube                 & 5.29\%  & \multicolumn{1}{c|}{21.77\%} & \multicolumn{1}{c|}{31.77\%} \\ \hline
$\langle\texttt{XLM}, Levenshtein\rangle$                                                       & YouTube                 & 18.24\%  & \multicolumn{1}{c|}{42.94\%} & \multicolumn{1}{c|}{47.65\%} \\ \hline
$\langle\texttt{XLM}, Phonetic\rangle$                                                          & YouTube                 & 7.06\%  & \multicolumn{1}{c|}{21.18\%}   & \multicolumn{1}{c|}{34.71\%} \\ \hline
$\langle\texttt{XLNet}, Levenshtein\rangle$      & YouTube                 & 10\%   & 28.83\%                      & 39.41\%   \\ \hline
$\langle\texttt{XLNet}, Phonetic\rangle$                                                        & YouTube                 & 0.59\%  & 17.65\%                      & 35.88\%                 \\ \hline
$\langle\texttt{DistilBERT}, Levenshtein\rangle$ & YouTube                 & 18.82\% & 36.47\%                      & 44.12\% \\ \hline
$\langle\texttt{DistilBERT}, Phonetic\rangle$                                                   & YouTube                 & 10\% & 22.94\%                      & 31.77\%                         \\ \hline
\cellcolor{blue!15}$\langle\texttt{Megatron}, Levenshtein\rangle$   & YouTube                 & \cellcolor{blue!15}28.24\% & 39.41\%                      & 46.47\%        \\ \hline
$\langle\texttt{Megatron}, Phonetic\rangle$                                                     & YouTube                 & 10.59\%  & 28.24\%                      & 31.77\%                     \\ \hline
\end{tabular}
\vspace{0.5cm}
\caption{{Performance on our benchmark data set $\mathcal{D}_\mathit{taboo}$.}}
\label{tab:ClozeResults}
\end{table}


\section{Conclusions and Discussions}

In this paper, we have found a disturbing result that commercial ASR systems may hallucinate taboo-words in video content for children. On a data set consisting of highly popular videos with worldwide consumption, we show that such hallucinations are far from occasional errors. We release a one-of-its-kind challenging data set of audio inputs where high-performance ASR systems have hallucinated taboo-words. We also show that some of these hallucinations can be corrected using language models.

Our work raises several important points to ponder.\\ 
\noindent\textbf{\emph{1. Which words are inappropriate for kids?}} Deciding on the set of inappropriate words for kids was one of the major design issues we ran into in this project. We considered several existing literature, published lexicons, and also drew from popular children's entertainment content. However, we felt that much needs to be done in reconciling the notion of inappropriateness and changing times. For example, we found that both $\mathcal{H}_1$ and $\mathcal{H}_2$ contain terms such as \texttt{gay} and \texttt{queer}. The field study that yielded $\mathcal{H}_2$ was conducted in early 1990s. Additionally, these words may or may not appear as abusive content based on the context it is present in. Since then, the continual struggle for LGBTQ+ rights and equality has made massive strides. Although queer studies is a developing field, \cite{CAMPO-ARIAS2010} demonstrates that a child’s age when they become aware of their sexual orientation varies, and it is possible that it could occur during childhood. In addition to this, children's attitudes toward queer people are also positively influenced by media exposure \cite{article}, but they can also vary due to cultural differences \cite{doi:10.1177/0022022111420146}. We thus strongly feel that these lexicons need revisiting from experts to set better ethical guidelines for kids' content reflective of modern times.\\ 

\textbf{\emph{2. Risks of black-box AI systems.}} Several recent lines of work have reported instances where state-of-the-art content-filtering systems got blindsided by unseen~\cite{DBLP:conf/aaai/SarkarK21} or adversarial content~\cite{grondahl2018all}. Recent studies have revealed that biases in large language models often influence toxic content generation in neural text generation models~\cite{DBLP:conf/emnlp/GehmanGSCS20}. In many such cases, we are dealing with opaque systems where it is impossible to know on what data these large systems are trained on, a risk aptly discussed in~\cite{10.1145/3442188.3445922}. At a philosophical level, we see our work making a small contribution in this growing discussion of responsible, inclusive, and trustworthy AI in the following key ways. First, we show that downstream AI  applications can introduce highly inappropriate taboo-words in kids' content originally not present in them. The benefits of these ASR systems are undeniable. That said, we cannot disregard the fact that such systems when applied to content with high visibility to a vulnerable and impressionable community needs rigorous checks and balances. 
Our findings, backed up with a challenging benchmark data set, is a small step towards that. Second, we do not know the distribution of kids' speech examples in the training data these opaque systems are trained on, nor do we know how well these data sets  represent ESL (English as Second Language) speakers. However, our analysis of hallucinations reveal that many of the errors were caused in presence of ESL speakers and kids. Our results thus potentially point to ways these data sets can be more inclusive. Finally, our work draws the attention of the community to form deeper understanding of intermediate risks in a chain of black-box systems, where one system's outputs are inputs to another.\\

\noindent\textbf{\emph{3. Mitigation strategies.}} In our cloze test experiments to fix some of these hallucinated taboo-words, we notice that language models often have a propensity towards predicting the taboo-word as the most likely completion. For instance, according to \texttt{BERT}, a constrained cloze test \emph{I love [MASK]} with candidate sets \{\texttt{porn}, \texttt{corn}\} yields \texttt{porn} as the likelier completion. In fact, 16.90\% (Google) and 14.95\% (Amazon) of top predictions by the Megatron model were taboo words. This indicates that although language models can bring in improvement, they alone cannot fix the problem as these models also possibly suffer from a similar issue of being trained on content largely meant for an adult audience as opposed to kids. 

While we observe limited success in fixing some of the hallucinated taboo-words, our experiments revealed potential avenues for improvement. We observe that many of the hallucinated audio inputs had visual signals that can be leveraged. For instance, in examples where \texttt{crab} is confused with \texttt{crap}, object recognition information can complement textual information to correct such mistakes. A multimodal method to robustify ASR systems could be a worthy future research challenge. Our experiments with language models produced a modest \textbf{p@1} improvement.  However, a better \textbf{p@10} performance indicates that a human-in-the-loop setting, coupled with suggestions from language models, especially given these contents are consumed by kids worldwide, can offer more safety to kids.\\   

\noindent\textbf{\emph{4. Integration challenges between YouTube Kids and general YouTube.}} YouTube Kids allows keyword-based search if parents (or guardians) enable it in the application. Of the five highly inappropriate taboo-words, \{\texttt{shit}, \texttt{fuck}, \texttt{crap}, \texttt{rape}, \texttt{ass}\}, we find that \texttt{rape}, \texttt{fuck}, and \texttt{shit} are not searchable through the kids app (understandably). We also find that most English language subtitles (including subtitles with many hallucinated taboo-words) are disabled on the kids app. However, as shown in Figure \ref{fig:example}, the same videos have subtitles enabled on general YouTube. It is unclear how often kids are only confined to the YouTube Kids app while watching videos and how frequently parents (or guardians) simply let them watch kids content from general YouTube. Our findings indicate a need for tighter integration between YouTube general and YouTube kids to be more vigilant about kids' safety.     



\section{Acknowledgement}
We acknowledge the support received from Srini Raju Centre for IT and the Networked Economy (SRITNE) at the Indian School of Business (ISB). We thank  Mallikarjuna Tupakula's help in the data collection and acknowledge Philip G. Bittenbender for his inputs in annotation. 

\bibliographystyle{unsrt}

\begin{thebibliography}{10}

\bibitem{papadamou2020disturbed}
Kostantinos Papadamou, Antonis Papasavva, Savvas Zannettou, Jeremy Blackburn,
  Nicolas Kourtellis, Ilias Leontiadis, Gianluca Stringhini, and Michael
  Sirivianos.
\newblock Disturbed youtube for kids: Characterizing and detecting
  inappropriate videos targeting young children.
\newblock In {\em Proceedings of the international AAAI conference on web and
  social media}, volume~14, pages 522--533, 2020.

\bibitem{han2020discovery}
Wenlin Han and Madhura Ansingkar.
\newblock Discovery of elsagate: Detection of sparse inappropriate content from
  kids videos.
\newblock In {\em 2020 Zooming Innovation in Consumer Technologies Conference
  (ZINC)}, pages 46--47. IEEE, 2020.

\bibitem{alghowinem2018safer}
Sharifa Alghowinem.
\newblock {A Safer YouTube Kids: An Extra Layer of Content Filtering Using
  Automated Multimodal Analysis}.
\newblock In {\em Proceedings of SAI Intelligent Systems Conference}, pages
  294--308. Springer, 2018.

\bibitem{riccardi1997spoken}
Giuseppe Riccardi, Allen~L Gorin, Andrej Ljolje, and Michael Riley.
\newblock A spoken language system for automated call routing.
\newblock In {\em 1997 IEEE International Conference on Acoustics, Speech, and
  Signal Processing}, volume~2, pages 1143--1146. IEEE, 1997.

\bibitem{ranchal2013using}
Rohit Ranchal, Teresa Taber-Doughty, Yiren Guo, Keith Bain, Heather Martin,
  J~Paul Robinson, and Bradley~S Duerstock.
\newblock Using speech recognition for real-time captioning and lecture
  transcription in the classroom.
\newblock {\em IEEE Transactions on Learning Technologies}, 6(4):299--311,
  2013.

\bibitem{mehrabani2015personalized}
Mahnoosh Mehrabani, Srinivas Bangalore, and Benjamin Stern.
\newblock Personalized speech recognition for internet of things.
\newblock In {\em 2015 IEEE 2nd World Forum on Internet of Things (WF-IoT)},
  pages 369--374. IEEE, 2015.

\bibitem{finley2018automated}
Gregory Finley, Erik Edwards, Amanda Robinson, Michael Brenndoerfer, Najmeh
  Sadoughi, James Fone, Nico Axtmann, Mark Miller, and David Suendermann-Oeft.
\newblock An automated medical scribe for documenting clinical encounters.
\newblock In {\em Proceedings of the 2018 Conference of the North American
  Chapter of the Association for Computational Linguistics: Demonstrations},
  pages 11--15, 2018.

\bibitem{jay1992cursing}
Timothy Jay.
\newblock {\em Cursing in america}, volume~10.
\newblock Philadelphia: John Benjamins, 1992.

\bibitem{vanderplank2016effects}
Robert Vanderplank.
\newblock ‘effects of’and ‘effects with’captions: how exactly does
  watching a tv programme with same-language subtitles make a difference to
  language learners?
\newblock {\em Language Teaching}, 49(2):235--250, 2016.

\bibitem{vanderplank2016captioned}
Robert Vanderplank.
\newblock {\em Captioned media in foreign language learning and teaching:
  Subtitles for the deaf and hard-of-hearing as tools for language learning}.
\newblock Springer, 2016.

\bibitem{wu2019advances}
Fei Wu, Leibny~Paola Garc{\'\i}a-Perera, Daniel Povey, and Sanjeev Khudanpur.
\newblock Advances in automatic speech recognition for child speech using
  factored time delay neural network.
\newblock In {\em Interspeech}, pages 1--5, 2019.

\bibitem{yeung2018difficulties}
Gary Yeung and Abeer Alwan.
\newblock On the difficulties of automatic speech recognition for
  kindergarten-aged children.
\newblock {\em Interspeech 2018}, 2018.

\bibitem{plantinga2019towards}
Peter Plantinga and Eric Fosler-Lussier.
\newblock Towards real-time mispronunciation detection in kids' speech.
\newblock In {\em 2019 IEEE Automatic Speech Recognition and Understanding
  Workshop (ASRU)}, pages 690--696. IEEE, 2019.

\bibitem{sawaf2012automatic}
Hassan Sawaf.
\newblock Automatic speech recognition and hybrid machine translation for
  high-quality closed-captioning and subtitling for video broadcast.
\newblock {\em Proceedings of Association for Machine Translation in the
  Americas--AMTA}, 14, 2012.

\bibitem{rajnoha2011asr}
Josef Rajnoha and Petr Poll{\'a}k.
\newblock Asr systems in noisy environment: Analysis and solutions for
  increasing noise robustness.
\newblock {\em Radioengineering}, 20(1):74--84, 2011.

\bibitem{benzeghiba2007automatic}
Mohamed Benzeghiba, Renato De~Mori, Olivier Deroo, Stephane Dupont, Teodora
  Erbes, Denis Jouvet, Luciano Fissore, Pietro Laface, Alfred Mertins,
  Christophe Ris, et~al.
\newblock Automatic speech recognition and speech variability: A review.
\newblock {\em Speech communication}, 49(10-11):763--786, 2007.

\bibitem{10.5555/2018192.2018228}
Kinfe~Tadesse Mengistu and Frank Rudzicz.
\newblock Comparing humans and automatic speech recognition systems in
  recognizing dysarthric speech.
\newblock In {\em Proceedings of the 24th Canadian Conference on Advances in
  Artificial Intelligence}, Canadian AI'11, page 291–300, Berlin, Heidelberg,
  2011. Springer-Verlag.

\bibitem{10.1145/2899475.2899478}
Yashesh Gaur, Walter~S. Lasecki, Florian Metze, and Jeffrey~P. Bigham.
\newblock The effects of automatic speech recognition quality on human
  transcription latency.
\newblock In {\em Proceedings of the 13th International Web for All
  Conference}, W4A '16, New York, NY, USA, 2016. Association for Computing
  Machinery.

\bibitem{arisoy2015bidirectional}
Ebru Arisoy, Abhinav Sethy, Bhuvana Ramabhadran, and Stanley Chen.
\newblock Bidirectional recurrent neural network language models for automatic
  speech recognition.
\newblock In {\em 2015 IEEE International Conference on Acoustics, Speech and
  Signal Processing (ICASSP)}, pages 5421--5425. IEEE, 2015.

\bibitem{chen2017investigating}
Xie Chen, Anton Ragni, Xunying Liu, and Mark~JF Gales.
\newblock Investigating bidirectional recurrent neural network language models
  for speech recognition.
\newblock In {\em Proceedings of Interspeech 2017}, pages 269--273.
  International Speech Communication Association (ISCA), 2017.

\bibitem{shin2019effective}
Joonbo Shin, Yoonhyung Lee, and Kyomin Jung.
\newblock Effective sentence scoring method using bert for speech recognition.
\newblock In {\em Asian Conference on Machine Learning}, pages 1081--1093.
  PMLR, 2019.

\bibitem{DBLP:journals/corr/abs-2103-14580}
Mahdi Namazifar, John Malik, Li~Erran Li, G{\"{o}}khan T{\"{u}}r, and Dilek
  Hakkani{-}T{\"{u}}r.
\newblock Correcting automated and manual speech transcription errors using
  warped language models.
\newblock {\em CoRR}, abs/2103.14580, 2021.

\bibitem{kim2019comparison}
Joshua~Y Kim, Chunfeng Liu, Rafael~A Calvo, Kathryn McCabe, Silas~CR Taylor,
  Bj{\"o}rn~W Schuller, and Kaihang Wu.
\newblock A comparison of online automatic speech recognition systems and the
  nonverbal responses to unintelligible speech.
\newblock {\em arXiv preprint arXiv:1904.12403}, 2019.

\bibitem{10.1145/3442188.3445922}
Emily~M. Bender, Timnit Gebru, Angelina McMillan-Major, and Shmargaret
  Shmitchell.
\newblock On the dangers of stochastic parrots: Can language models be too big?
\newblock In {\em Proceedings of the 2021 ACM Conference on Fairness,
  Accountability, and Transparency}, FAccT '21, page 610–623. Association for
  Computing Machinery, 2021.

\bibitem{DBLP:conf/emnlp/GehmanGSCS20}
Samuel Gehman, Suchin Gururangan, Maarten Sap, Yejin Choi, and Noah~A. Smith.
\newblock Realtoxicityprompts: Evaluating neural toxic degeneration in language
  models.
\newblock In Trevor Cohn, Yulan He, and Yang Liu, editors, {\em Findings of the
  Association for Computational Linguistics: {EMNLP} 2020, Online Event, 16-20
  November 2020}, volume {EMNLP} 2020 of {\em Findings of {ACL}}, pages
  3356--3369. Association for Computational Linguistics, 2020.

\bibitem{sutton1978psychosexual}
Brian Sutton-Smith and David~M Abrams.
\newblock Psychosexual material in the stories told by children: The fucker.
\newblock {\em Archives of Sexual Behavior}, 7(6):521--543, 1978.

\bibitem{jay2013child}
Kristin~L Jay and Timothy~B Jay.
\newblock A child's garden of curses: A gender, historical, and age-related
  evaluation of the taboo lexicon.
\newblock {\em The American Journal of Psychology}, 126(4):459--475, 2013.

\bibitem{devlin2018bert}
Jacob Devlin, Ming-Wei Chang, Kenton Lee, and Kristina Toutanova.
\newblock {BERT}: Pre-training of deep bidirectional transformers for language
  understanding.
\newblock In {\em Proceedings of the 2019 Conference of the North {A}merican
  Chapter of the Association for Computational Linguistics: Human Language
  Technologies, Volume 1 (Long and Short Papers)}, pages 4171--4186, June 2019.

\bibitem{chi2021xlme}
Zewen Chi, Shaohan Huang, Li~Dong, Shuming Ma, Saksham Singhal, Payal Bajaj,
  Xia Song, and Furu Wei.
\newblock Xlm-e: Cross-lingual language model pre-training via electra, 2021.

\bibitem{DBLP:journals/corr/abs-1910-01108}
Victor Sanh, Lysandre Debut, Julien Chaumond, and Thomas Wolf.
\newblock Distilbert, a distilled version of {BERT:} smaller, faster, cheaper
  and lighter.
\newblock {\em CoRR}, abs/1910.01108, 2019.

\bibitem{DBLP:journals/corr/abs-1906-08237}
Zhilin Yang, Zihang Dai, Yiming Yang, Jaime~G. Carbonell, Ruslan Salakhutdinov,
  and Quoc~V. Le.
\newblock Xlnet: Generalized autoregressive pretraining for language
  understanding.
\newblock {\em CoRR}, abs/1906.08237, 2019.

\bibitem{DBLP:journals/corr/abs-1909-08053}
Mohammad Shoeybi, Mostofa Patwary, Raul Puri, Patrick LeGresley, Jared Casper,
  and Bryan Catanzaro.
\newblock Megatron-lm: Training multi-billion parameter language models using
  model parallelism.
\newblock {\em CoRR}, abs/1909.08053, 2019.

\bibitem{taylor1953cloze}
Wilson~L Taylor.
\newblock “cloze procedure”: A new tool for measuring readability.
\newblock {\em Journalism quarterly}, 30(4):415--433, 1953.

\bibitem{petroni-etal-2019-language}
Fabio Petroni, Tim Rockt{\"a}schel, Sebastian Riedel, Patrick Lewis, Anton
  Bakhtin, Yuxiang Wu, and Alexander Miller.
\newblock Language models as knowledge bases?
\newblock In {\em Proceedings of the 2019 Conference on Empirical Methods in
  Natural Language Processing and the 9th International Joint Conference on
  Natural Language Processing (EMNLP-IJCNLP)}, pages 2463--2473, Hong Kong,
  China, November 2019. Association for Computational Linguistics.

\bibitem{electionKhudaBukhsh}
Shriphani Palakodety, Ashiqur~R. KhudaBukhsh, and Jaime~G. Carbonell.
\newblock Mining insights from large-scale corpora using fine-tuned language
  models.
\newblock In {\em {ECAI} 2020 - 24th European Conference on Artificial
  Intelligence}, volume 325 of {\em Frontiers in Artificial Intelligence and
  Applications}, pages 1890--1897. {IOS} Press, 2020.

\bibitem{bert-score}
Tianyi Zhang*, Varsha Kishore*, Felix Wu*, Kilian~Q. Weinberger, and Yoav
  Artzi.
\newblock Bertscore: Evaluating text generation with bert.
\newblock In {\em International Conference on Learning Representations}, 2020.

\bibitem{sarkar-etal-2020-non}
Rupak Sarkar, Sayantan Mahinder, and Ashiqur KhudaBukhsh.
\newblock The non-native speaker aspect: {I}ndian {E}nglish in social media.
\newblock In {\em Proceedings of the Sixth Workshop on Noisy User-generated
  Text (W-NUT 2020)}, pages 61--70, Online, November 2020. Association for
  Computational Linguistics.

\bibitem{CAMPO-ARIAS2010}
Adalberto~Fort{\'u}n Campo.
\newblock Essential aspects and practical implications of sexual identity.
\newblock {\em Colombia Medica}, 41:179--185, 2010.

\bibitem{article}
Shiqi Zhang, Shuohan Feng, and Zhimin Shen.
\newblock How do background factors influence children’s attitudes toward
  gays and lesbians?
\newblock {\em Psychology}, 10:1572--1594, 01 2019.

\bibitem{doi:10.1177/0022022111420146}
Henny M.~W. Bos, Charles Picavet, and Theo G.~M. Sandfort.
\newblock Ethnicity, gender socialization, and children’s attitudes toward
  gay men and lesbian women.
\newblock {\em Journal of Cross-Cultural Psychology}, 43(7):1082--1094, 2012.
\newblock PMID: 23162164.

\bibitem{DBLP:conf/aaai/SarkarK21}
Rupak Sarkar and Ashiqur~R. KhudaBukhsh.
\newblock Are chess discussions racist? an adversarial hate speech data set
  (student abstract).
\newblock In {\em Thirty-Fifth {AAAI} Conference on Artificial Intelligence,
  {AAAI} 2021}, pages 15881--15882. {AAAI} Press, 2021.

\bibitem{grondahl2018all}
Tommi Gr{\"o}ndahl, Luca Pajola, Mika Juuti, Mauro Conti, and N~Asokan.
\newblock All you need is" love" evading hate speech detection.
\newblock In {\em Proceedings of the 11th ACM workshop on artificial
  intelligence and security}, pages 2--12, 2018.

\end{thebibliography}

\end{document}